\documentstyle[sprocl]{article}

\newcommand{\dN}{\mbox{{\rm I \hspace{-0.865em} N}}} 
\newcommand{\CM}{{\cal M}}

\newcommand{\ba}{\begin{array}}
\newcommand{\ea}{\end{array}}

\newcommand{\beq}{\begin{equation}}
\newcommand{\eeq}{\end{equation}}
\newcommand{\bea}{\begin{eqnarray}}
\newcommand{\eea}{\end{eqnarray}}
\newcommand{\beal}{\setcounter{letter}{1} \begin{eqnarray}}
\newcommand{\eeal}{\addtocounter{equation}{1} \end{eqnarray}}

\newcommand{\req}[1]{Eq.(\ref{#1})}

\newcommand{\larrow}{\,\,\,\,\hbox to 30pt{\rightarrowfill}
\,\,\,\,}
\newcommand{\slarrow}{\,\,\,\hbox to 20pt{\rightarrowfill}
\,\,\,}

\begin{document}

\title 
{STATISTICAL MECHANICAL ENTROPY OF TWO-DIMENSIONAL BLACK HOLES}
\author
{J. Gegenberg} 
\address {Dept. of Mathematics and Statistics,
University of New Brunswick\\
Fredericton, New Brunswick E3B 5A3  
Canada}
\author{G. Kunstatter}
\address{Dept. of Physics and Winnipeg Institute of
Theoretical Physics, University of Winnipeg\\
Winnipeg, Manitoba R3B 2E9
Canada} 
\author{T.Strobl}
\address{Institut f\"ur Theoretische Physik, 
RWTH-Aachen\\
Sommerfeldstr. 26-28, D52056 Aachen, 
Germany}
\maketitle\abstracts{
We calculate the statistical mechanical entropy associated with
boundary 
terms in the two-dimensional Euclidean black holes in
deSitter 
gravity.}

\noindent PITHA - 96/22\\
\noindent UNB Technical Report 96-01

\section{Introduction}

Although the final word has certainly not
been written, several calculations have been performed
which relate the Bekenstein-Hawking thermodynamic entropy 
\cite{bek} 
to the degeneracy of microscopic states.  
For Carlip \cite{carlip} 
and Balachandran and his group \cite{bal},
the microstates originate from treating the event horizon as the
boundary of the spacetime.  In particular, degrees of freedom which
are pure gauge in the interior of spacetime become dynamical on the
boundary.

\par
In this talk, we apply Carlip's program to an even lower dimensional
black hole -- namely that 
occurring as a solution of the `constant
curvature' (CC) theory  
of gravity in two-dimensional spacetime with a
cosmological constant.\cite{jt} 
 The field equations for this theory
include $R=2k$, for Ricci scalar $R$ and cosmological constant $k\neq
0$.  We work with a closely related BF type topological gauge
theory.\cite{it}  We will show that when the latter theory is constructed on
a manifold with a boundary that corresponds to the bifurcation 
two-sphere of a black hole of fixed mass $M$, and provided the gauge
group is analytically continued to $SU(2)$ (or, equivalently, if we
consider the Euclidean CC-model,) 
the surface term gives rise to a non-trivial mechanical system
equivalent to two harmonic oscillators with fixed energy. This system
can be quantized only if the mass of the black hole is quantized. The
resulting degeneracy of quantum states on the boundary is proportional
to the square root of the black hole mass, and hence the entropy is
given by $\log{\sqrt{M}}$.

\section{The 2-D Gravity Model}
The CC theory is closely related to a gauge theory \cite{it} 
\beq
S[\phi, A]=Tr \int_\CM \phi F(A)- Tr \, \oint_{\partial \CM} \phi A   \, ,
\label{bf}
\eeq where $F(A)=dA+A\wedge A$ is the curvature of $A$, a Lie-algebra
valued connection on a principal bundle over $\CM$ and $\phi$ is a
scalar field over $\CM$ which also takes its values in the same Lie
algebra.  In order to describe the Euclidean  
CC theory with cosmological constant $k$, the appropriate Lie algebra
is $so(2,1)$ for negative
cosmological constant and $su(2)$ for positive $k$, while for the
Minkowskian theory it is  $so(2,1)$ irrespective of the sign of
$k$.\cite{it} 
The gauge group thus coincides with the isometry group of the on-shell
metric. The equivalence with the CC gravity model is achieved by the
identification  $A=A^iT_i=e^aT_a+\omega T_3$, where $a,b,..=1,2$ and 
$e^a$, $\omega$ are Zweibein and spin connection, respectively.
The surface term is required in \req{bf}, if the spacetime
is a manifold-with-boundary such that 
the field $\phi$ is fixed on the boundary.
\par
We will henceforth suppose that the spacetime metric given by the dyad
$e^a$ has positive-definite signature and that $k$ is positive. Thus
the appropriate gauge group is $SU(2)$.  The black holes are to be
identified with the manifold-with-boundary $\CM$ with interior the
spherical cap and with boundary $\partial \CM$ 
the embedded circle.  The metric
is given locally by \beq
ds^2=\left(M-x^2/\ell^2\right)d\theta^2+\left(M-x^2/\ell^2\right)^{-1}dx^2,
\eeq where $M$ is an invariant dimensionless parameter defined by
$M=Tr\phi^2 \equiv \phi^a\phi_a+ \phi^3\phi^3$, which may be identified
with the mass of the black hole.
This is the dimensionally reduced
Euclideanized BTZ black hole.\cite{btz}  The boundary in these
coordinates is given by $x = x_0 > \sqrt{M}\ell$.

\par
It turns out 
that on-shell the fields $\phi^a$ may be identified  with the 
Killing vector $k$ of the black hole solution:  $\phi^a=e^a_\mu k^\mu.  $
Thus to impose  boundary conditions
corresponding to the bifurcation surface of the event horizon we 
will require $\phi^a=0$ and, as a consequence, $\phi_3=\sqrt{M}$.

\par
It is important to note that the surface term in \req{bf} is not gauge
invariant under $SU(2)$ gauge transformations.  The gauge
transformations generate spacetime diffeomorphisms \cite{it}, and
since some diffeomorphisms shift the location of the boundary, one
expects that some gauge degrees of freedom in the interior of $\CM$ are
dynamical degrees of freedom on the boundary.\cite{carlip} Hence, as
in Carlip's discussion of the three-dimensional case \cite{carlip}, we
parametrize the fields $\phi,A$ in terms of gauge-fixed fields
$\bar\phi,\bar A$ and an element $g\in \, SU(2)$: \beq \phi=
g^{-1}\bar\phi g \, ;\,\, A=g^{-1}\bar Ag+g^{-1}dg.\label{param} \eeq
This yields  the boundary action \beq S[g]=-Tr\oint_{\partial \CM} 
\bar\phi dg\,g^{-1}
\label{baction} \eeq describing gauge modes that have become physical
due to the imposed black hole boundary conditions.  In the next
section we will quantize the theory with action \req{baction}.

\section{Quantum Mechanics on the Boundary of Spacetime} 
As discussed above, we choose $\phi=\bar\phi$ to be:
$\bar\phi=\sqrt{M} \, T_3,
$  where $T_3$ is the constant diagonal 
 $2\times 2$ {\it su}(2) matrix.   
This is a solution of the equations of motion  
on the boundary $\partial \CM$, in the limit as $x_0 \to\sqrt{M}\ell$. 

\par

The theory \req{baction} is a classical {\it mechanical} one: it
describes a particle moving  on the homogenous space 
$SU(2)/U(1)$ (equivalently, on the coadjoint orbit of $su(2)$)
with periodic time $t$.  This is a result of
the fact that a $g$ in the diagonal subgroup $U(1)$ of $SU(2)$ does not
contribute to the action -- such terms are total divergences.  In the
following we quantize this point particle system in an elementary way,
without resort to more elaborate procedures such as, e.g.,  geometric
quantization.  We parametrize $g$ by \beq g=\left[\ba{clcr}
z_1\,\,&z_2\\-\bar z_2\,\,&\bar z_1\ea\right], \eeq where \beq
z_1={1\over2}\left({q_1\over\sqrt{M}}-ip_1\right) \, ,\,\,
z_2={1\over2}\left(p_2+i{q_2\over\sqrt{M}}\right), \eeq with the $p$'s
and $q$'s real.  Then the action \req{baction} becomes simply
$S[q,p]=\oint_{\partial \CM} 
dt\left(p_1\dot q_1+p_2\dot q_2\right)$. However,
since $g\in$ $SU(2)$, there is a first-class constraint \beq \mid
z_1\mid^2+\mid z_2\mid^2 \, \, \, \equiv  \, \,
(p_1)^2 + (p_2)^2 + {1\over M}\left((q_1)^2
+ (q_2)^2\right) \, \approx \, 1.\label{constraint} \eeq 
Implementation of this constraint on the quantum level (plus
normalizability of the physical wave functions) maps the system to the
quantum system of two identical harmonic oscillators with fundamental
frequency $\omega\propto 1/\sqrt{M}$ and total energy constrained to
be 1. This obviously is consistent only if $M$ is discrete
(quantized). Counting of the degeneracy of the energy eigenvalue $1 =
\hbar \omega (n + 1/2)$, $n \in \, \dN$, is elementary now. It is easy
to see that it yields the announced result of an  entropy $\sim 
\log{\sqrt{M}}$. We observe in parenthesis  
that the period  associated to the above oscillator frequency, 
 $2\pi/\omega=2\pi\sqrt{M}$, produces 
 the correct periodicity in  Euclidean time   for a 
black hole with Hawking temperature $T_H= \sqrt{M}/2\pi$.

It is fairly well-known \cite{ent} that the Bekenstein-Hawking entropy of
the CC black holes goes as $\sqrt{M}$.  Hence 
there is an insufficient degeneracy of states on the
boundary to account for the Bekenstein-Hawking entropy in the case of
the 2d Euclidean CC black holes. 

Seemingly this changes into the opposite when
looking at the Minkowskian CC theory: There the non-compactness of the
gauge group gives rise to an {\em infinite\/} number of quantum states
living at the boundary. 
In the framework of Poisson $\sigma$-models \cite{strobl},
furthermore, much of 
the above investigation may be generalized to generic 
2d dilaton gravity theories. Remarkably, in many cases 
this yields infinitely many boundary states even for Euclidean 
signature black holes. 
Work on this is in progress and shall be reported on 
elsewhere.

\noindent{\bf Acknowledgment} The authors would like to thank Steve
Carlip for useful discussions.  This work was supported in part by
the Natural Sciences and Engineering Research Council of Canada as
well as by the Austrian FWF Project 10.221-PHY.


\begin{thebibliography}{99}

\bibitem{bek}  J.D. Bekenstein, Phys. Rev. {\bf D7}, 2333 (1973); S.W. Hawking, 
Nature {\bf248}, 30 (1974).

\bibitem{carlip}  S. Carlip, 
Phys. Rev. {\bf D51}, 632 (1995).

\bibitem{bal}  A.P. Balachandran, L. Chandar and A. Momen, ``Edge States 
in Gravity and Black Hole Physics", gr-qc/9412019 (1994); ``Edge States in Canonical Gravity", gr-qc/9506006 (1995).

\bibitem{jt}  B.M. Barbashov, V.V. Nesterenko and A.M. Chervjakov,
Theor. Math. Phys. {\bf 40}, 15 (1979);  R. Jackiw in {\it Quantum
Theory of Gravity}, ed. by S.  Christensen, Adam Hilger, Bristol
(1984);  C. Teitelboim in {\it Quantum Theory of Gravity}, ed. by S.
Christensen, Adam Hilger, Bristol (1984).

\bibitem{it}  T. Fukuyama and K. Kamimura, Phys. Lett. B{\bf 160}, 259 
(1985);  K. Isler and C. Trugenberger, Phys. Rev. Lett. {\bf 63}, 834 
(1989);  A. Chamseddine and D. Wyler, Phys. Lett. B{\bf 228}, 75
(1989); cf.\ also P. Schaller and T. Strobl, Phys. Lett. B{\bf 337},
266. 

\bibitem{btz}  M. Banados,  C. Teitelboim and J. Zanelli, 
Phys. Rev. Lett. {\bf69}, 1849 (1992); M. Banados,  M. Henneaux,
C. Teitelboim and  J. Zanelli, Phys. Rev. {\bf D48}, 1506 (1993).

\bibitem{ent}  J. Gegenberg, G. Kunstatter and D.Louis-Martinez, Phys. 
Rev. {\bf D51}, 1781 (1995).

\bibitem{strobl} P. Schaller and T. Strobl, Mod. Phys. Letts. {\bf A9}, 
3129 (1994);  ``A Brief Introduction to Poisson $\sigma$-Models", 
hep-th/9507020 (1995). 

\end{thebibliography}
\end{document}